\documentclass[conference, 10pt]{IEEEtran}
\usepackage{hyperref}
\usepackage[cmex10]{amsmath}

\usepackage[noadjust]{cite}

\usepackage[pdftex]{graphicx}
\graphicspath{{./figures/}}
\DeclareGraphicsExtensions{.pdf,.jpeg,.png}

\usepackage{standalone}

\ifCLASSOPTIONcompsoc
  \usepackage[caption=false,font=normalsize,labelfont=sf,textfont=sf]{subfig}
\else
  \usepackage[caption=false,font=footnotesize]{subfig}
\fi

\usepackage{tikz}
\newcommand\copyrighttext{%
  \footnotesize \textcopyright 2015 IEEE. Personal use of this material is permitted. Permission from IEEE must be obtained for all other uses, in any current or future media, including reprinting/republishing this material for advertising or promotional purposes, creating new collective works, for resale or redistribution to servers or lists, or reuse of any copyrighted component of this work in other works. DOI: \href{http://dx.doi.org/10.1109/GLOCOMW.2015.7414047}{10.1109/GLOCOMW.2015.7414047}}
\newcommand\copyrightnotice{%
\begin{tikzpicture}[remember picture,overlay]
\node[anchor=south,yshift=10pt] at (current page.south) {\fbox{\parbox{\dimexpr\textwidth-\fboxsep-\fboxrule\relax}{\copyrighttext}}};
\end{tikzpicture}%
}
\begin{document}
\title{Applying Bag of System Calls for Anomalous Behavior Detection of Applications in Linux Containers}

\author{\IEEEauthorblockN{Amr S. Abed}
\IEEEauthorblockA{Department of Electrical \& Computer Engineering\\
Virginia Tech, Blacksburg, VA\\
amrabed@vt.edu}
\and
\IEEEauthorblockN{T. Charles Clancy, David S. Levy}
\IEEEauthorblockA{Hume Center for National Security \& Technology\\
Virginia Tech, Arlington, VA\\
\{tcc, dslevy\}@vt.edu}}
\maketitle
\copyrightnotice
\begin{abstract}
In this paper, we present the results of using bags of system calls for learning the behavior of Linux containers for use in anomaly-detection based intrusion detection system. By using system calls of the containers monitored from the host kernel for anomaly detection, the system does not require any prior knowledge of the container nature, neither does it require altering the container or the host kernel.
\end{abstract}
\IEEEpeerreviewmaketitle
\section{Introduction}
Linux containers are computing environments apportioned and managed by a host kernel. Each container typically runs a single application that is isolated from the rest of the operating system. A Linux container provides a runtime environment for applications and individual collections of binaries and required libraries. Namespaces are used to assign customized views, or permissions, applicable to its needed resource environment. Linux containers typically communicate with the host kernel via system calls. 

By monitoring the system calls between the container and the host kernel, one can learn the behavior of the container in order to detect any change of behavior, which may reflect an intrusion attempt against the container. 

One of the basic approaches to anomaly detection using system calls is the \textit{Bag of System Calls} (BoSC) technique. The BoSC technique is a frequency-based anomaly detection technique, that was first introduced by Kang et al. in 2005~\cite{fuller2005}. Kang et al. define the bag of system call as an ordered list $<c_1, c_2, \dots, c_n>$, where $n$ is the total number of distinct system calls, and $c_i$ is the number of occurrences of the system call, $s_i$, in the given input sequence. BoSC has been used for anomaly detection at the process level~\cite{fuller2005} and at the level of virtual machines (VMs)~\cite{mutz2006}\cite{alarifi2012}\cite{alarifi2013}, and has shown promising results.  

The fewer number of processes in a container, as compared to VM, results in reduced complexity. The reduced complexity gives the potential for the BoSC technique to have high detection accuracy with a marginal impact on system performance when applied to anomaly detection in containers. 

In this paper, we study the feasibility of applying the BoSC to passively detect attacks against containers. The technique used is similar to the one introduced by~\cite{alarifi2012}. 
We show that a frequency-based technique is sufficient for detecting abnormality in container behavior. 

The rest of this paper is organized as follows. 
Section~\ref{sec_overview} provides an overview of the system.
Section~\ref{sec_experiments} describes the experimental design.
Section~\ref{sec_results} discusses the results of the experiments.
Section~\ref{sec_related} gives a brief summary of related work.
Section~\ref{sec_conclusion} concludes with summary and future work.
\section{System Overview}
\label{sec_overview}
In this paper, we use a technique similar to the one described in~\cite{alarifi2012} applied to Linux containers for intrusion detection. The technique combines the sliding window technique~\cite{forrest1996} with the bag of system calls technique~\cite{fuller2005} as described below. 

The system employs a background service running on the host kernel to monitor system calls between any Docker containers and the host Kernel. Upon start of a container, the service uses the Linux \texttt{strace} tool to trace all system calls issued by the container to the host kernel. The \texttt{strace} command reports system calls with their originating process ID, arguments, and return values. A table of all distinct system calls in the trace is also reported at the end of the trace along with the total number of occurrences. 

The full trace, and the count table, are stored into a log file that is processed offline and used to learn the container behavior after the container terminates. At this point, we are not performing any real-time behavior learning or anomaly detection. Therefore, dealing with the whole trace of the container offline is sufficient for our proof-of-concept purposes. However, for future purposes, where behavior learning and anomaly detection is to be achieved in real time (in which case the full trace would not be available), the learning algorithm applied would slightly differ from the one described here. However, the same underlying concepts will continue to apply.

The generated log file is then processed to create two files, namely syscall-list file and trace file. The syscall-list file holds a list of distinct system calls sorted by the number of occurrences. The trace file holds the full list of system calls as collected by \texttt{strace} after trimming off arguments, return values, and process IDs. The count file is used to create an syscall-index lookup table. 
The trace file is used to train the classifier, as described below.

The system reads the trace file epoch by epoch. For each epoch, a sliding window of size $10$ is moved over the system calls of the current epoch, counting the number of occurrences of each distinct system call in the current window, and producing a bag of system calls. As mentioned earlier, a bag of system calls is an array $<c_1, c_2, \dots, c_{n_s}>$ where $c_i$ is the number of occurrences of system call, $s_i$, in the current window, and $n_s$ is the total number of distinct system calls. When a new occurrence of a system call is encountered, the application retrieves the index of the system call from the syscall-index lookup table, and updates the corresponding index of the BoSC. For a window size of 10, the sum of all entries of the array equals $10$, i.e. $\sum_{i=1}^{n_s}{c_i}=10$. A sample BoSC is shown below for $n_s=20$ and sequence size of $10$.
\begin{verbatim}
[0,1,0,2,0,0,0,0,1,0,4,0,0,0,0,0,1,0,0,1]
\end{verbatim}

 If the current BoSC already exists in the normal-behavior database, its frequency is incremented by 1. Otherwise, the new BoSC is added to the database with initial frequency of $1$. 
The training is complete when all expected normal behavior patterns are applied to the system.

To detect whether an input trace is anomalous, the trace is read in epochs, and for each epoch, a sliding window is used to check if the current BoSC is present in the database of normal behavior database. If a BoSC is not present in the database, a mismatch is declared. The trace is declared anomalous if the number of mismatches within one epoch exceeds a certain threshold.
\section{Experimental Design}
\label{sec_experiments}
\subsection{Environment setup}
For our experiments, we are using a Docker container running on a Ubuntu Server 14.04 host operating system. The docker image we used for creating the container is the official \texttt{mysql} Docker image, which is basically a Docker image with MySQL 5.6 installed on a Debian operating system. 

On container start, the container automatically creates a default database, adds users defined by the environment variables passed to the container, and then starts listening for connections. Docker maps the MySQL port from the container to some custom port on the host.

Since there is no dataset available that contains system calls collected from containers, we needed to create our own datasets for both normal and anomalous behavior. For that, we created a container from the \texttt{mysql} Docker image. A normal-behavior work load was initially applied to the container, before it got ``attacked" using a penetration testing tool. 
\subsection{Generating normal dataset}
For generating normal-behavior dataset, we used \texttt{mysqlslap}~\cite{mysqlslap}; a program that emulates client load for a MySQL server. The tool has command-line options that allow the user to select the level of concurrency, and the number of iterations to run the load test. In addition, it gives the user the option to customize the created database, e.g. by specifying the number of \texttt{varcher} and/or \texttt{int} columns to use when creating the database. Moreover, the user can select the number of insertions and queries to perform on the database. 
The tool runs on the host kernel, and communicates with the MySQL server running on the container. 
\subsection{Simulating malicious attack}
To simulate an attack on the container, we used \texttt{sqlmap}~\cite{sqlmap}; an automatic SQL injection tool normally used for penetration testing purposes. In our experiment, we are using it to generate malicious-behavior dataset by attacking the MySQL database created on the container. Similarly, the \texttt{sqlmap} tool runs on the host kernel, and communicates with the attacked database through the Docker proxy. 

\subsection{Collecting and pre-processing data}
A background service, running on the host kernel, automatically detects any newly started Docker container, and traces system calls of the new container using the Linux \texttt{strace} tool. 

The service relies on the Docker command \texttt{events} to signal the service whenever a new container is started on the host kernel. Upon detection of the new container, the service starts to trace all processes running,  on container start, within the control group (cgroup) of the container. The service also traces any forked child processes by using the \texttt{-F} option of the \texttt{strace} tool. 

The behavior file is then passed to an extraction tool to split the output into two files, namely the count file and the trace file. The count file holds a list of distinct system calls sorted based on number of occurrences during the trace, while the trace file holds the full list of system calls as collected by \texttt{strace} after removing arguments, return values, and process number information.

\subsection{Modeling normal behavior}
We have implemented a Java application to learn the normal behavior of the container, i.e. train the classifier, using the detection technique described in section~\ref{sec_overview}. 

The application starts by building a syscall-index hash map from the count file. The hash map stores distinct system calls as the key, and a corresponding index as the value. A system call that appears in the whole trace less than the total number of distinct system calls is stored in the map as ``other". Using ``other" for relatively rarely-used system calls saves space, memory, and computation time, as described in~\cite{alarifi2012}. 

The application then reads the trace file in epochs. Each epoch updates the normal-behavior database. The normal-behavior database is another hash map with the BoSC as the key and the frequency of the bag as the value. If the current bag already exists in the database, the frequency value is incremented. Otherwise, a new entry is added to the database. 

For each epoch, the application uses the sliding window technique to read sequences of system calls from the trace file, with each sequence is of size 10. A bag of system calls is then created by counting the frequency of each distinct system call within the current window. The created bag of system calls is a frequency array of size $n_s$, where $n_s$ is the number of distinct system calls. When a new occurrence of a system call is encountered, the application retrieves the index of the system call from the syscall-index hash map, and the corresponding index of the frequency array is updated. The new BoSC is then added to the normal-behavior database.

After running each epoch $k$, where $k>1$,  the database is compared to a snapshot of the database before the epoch, and an array of BoSC frequency changes, $C_k$, is calculated. The cosine similarity metric is then used to find the similarity between $C_k$ and $C_{k-1}$ as shown in the equation below. Two vectors are identical if $\cos{(\theta)}$ is equal to $1$. 
\begin{align*}
\label{eq_cos}
\cos{(\theta_k)} &=\frac{C_k\cdot C_{k-1}}{\|C_k\|\|C_{k-1}\|} \\
&= \frac{\sum_{i=1}^{n_k}{C_k[i]C_{k-1}[i]}}{\sqrt{\sum_{i=1}^{n_k}{C_k[i]^2}}\sqrt{\sum_{i=1}^{n_{k-1}}{C_{k-1}[i]^2}}}
\end{align*}
where $C_k[i]$ is the $i^{th}$ entry of the $C_k$ array, and $n_k$ is the total number of database entries after epoch $k$.

The stopping condition for the training process is that the similarity metric is greater than or equal to certain threshold, $T_t$, for two consecutive epochs.

\subsection{Detecting anomaly}
The generated normal-behavior database is then applied to the post-attack trace of the container for anomaly detection. For each epoch, the sliding window technique is similarly used to check if the current BoSC is present in the database of normal behavior. A mismatch is declared whenever a BoSC is not present in the database. If the number of mismatches exceeds a certain threshold, $T_d$, within one epoch, an anomaly signal is raised. 
\section{Discussion}
\label{sec_results}
\subsection{Parameter Selection}
A number of parameters were intuitively selected when implementing our system. A list of this parameters is given below:
\begin{itemize}
\item ``Other" Threshold ($T_o$): The number of occurrences of a system call in a trace in order to be marked as ``other". This parameter is currently set to the total number of distinct system calls, i.e. the size of the syscall-index map.
\item Epoch Size ($S$): The total number of system calls in one epoch. It was noticed that the ratio between the epoch size and size of the input trace highly affects when and whether the training succeeds.
\item Sequence Size: A sequence size of $6$ or $10$ is usually recommended when using sliding-window techniques for better performance~\cite{forrest1996}\cite{forrest1999}\cite{hofmeyr1998}. Here, we are using $10$ since it was already shown for a similar work that size $10$ gives better performance than size $6$ without dramatically affecting the efficiency of the algorithm~\cite{alarifi2012}.
\item Training Threshold ($T_t$): The cosine-similarity value above which we declare two frequency-shift vectors as similar, currently $0.99$ (selected by experiment).
\item Detection Threshold ($T_d$): The number of detected mismatches before raising an anomaly signal, currently set to $10\%$ of the epoch size. 
\end{itemize}

Investigating the correlation between the values of the above parameters and the detection speed and accuracy, and formulating a way for selecting their optimal values, is left as future work.
\subsection{Classifier evaluation}
To evaluate the performance of the classifier, we are using the true positive rate (TPR) and false positive rate (FPR) metrics, defined as follows: 
 \begin{equation}
TPR=N_{tp}/N_{malicious}
\end{equation}
\begin{equation}
FPR=N_{fp}/N_{normal}
\end{equation}
where $N_{normal}$ and $N_{malicious}$ are the total number of normal and malicious sequences, respectively, and $N_{tp}$ and $N_{fp}$ are the number of true positives and false positives, respectively.
\subsection{Results}
We applied the described technique to a trace of $5,285,211$ system calls, using an epoch size ($S$) of $5000$. The trainer completed training after $37$ epochs. The number of distinct system calls ($n_s$) was $42$, and the size of the normal behavior database was $10809$ entries. 

The malicious data created a strong anomaly signal with an average of $2862$ mismatches per epoch, while the normal data had an average of $16$ mismatches per epoch. For $T_t=0.99$ and $T_d=0.1S$, the true positive rate is $100\%$ and the false positive rate is $0.58\%$. 

\begin{figure}
\centering
\subfloat[Accuracy]{\includegraphics[width=2.5in]{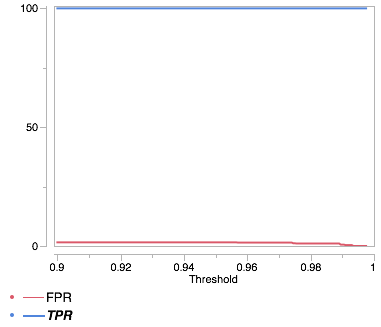}}
\hfil
\subfloat[Training Time]{\includegraphics[width=2.5in]{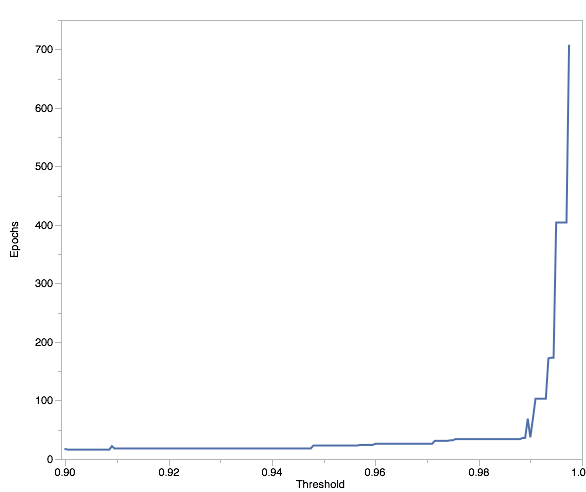}}
\caption{Effect of changing $T_t$ on accuracy and training time}
\label{fig_similarity}
\end{figure}

Figure~\ref{fig_similarity} shows the tradeoff between learning speed and accuracy. For instance, setting the training threshold ($T_t$) to $0.9935$ drops the FPR to $0\%$ at the expense of raising the number of epochs needed for training to $172$ epochs. 

For $T_t=0.99$ and $T_d=0.1S$, we needed a total of $185,000$ system calls to train the classifier. Applying the technique to VMs, the training process needed 5 epochs, each of $11,743,281$ system calls, to complete~\cite{alarifi2012}. Hence, it can be seen that the reduced complexity of the container resulted in a more efficient intrusion detection system as anticipated. 
\section{Related Work}
\label{sec_related}
Alarifi and Wolthusen used system calls for implementing a host-based intrusion detection for virtual machines residing in a multi-tenancy Infrastructure-as-a-service (IaaS) environment. They dealt with the VM as a single process, despite the numerous processes running inside it, and monitored system calls between the VM and the host operating system~\cite{alarifi2012}~\cite{alarifi2013}. 

In~\cite{alarifi2012}, they used the BoSC technique in combination with the sliding window technique for anomaly detection. In their technique, they read the input trace epoch by epoch. For each epoch, a sliding window of size $k$ moves over the system calls of each epoch, adding bags of system calls to the normal behavior database. The normal behavior database holds frequencies of bags of system calls. After building the normal-behavior database, i.e. training their classifier, an epoch is declared anomalous if the change in BoSC frequencies during that epoch exceeds certain threshold. For a sliding window of size $10$, their technique gave $100\%$ accuracy, with $100\%$ detection rate, and $0\%$ false positive rate. 

In~\cite{alarifi2013}, Alarifi and Wolthusen applied HMM for learning sequences of system calls for short-lived virtual machines. They based their decision on the conclusion from~\cite{forrest1999} that ``HMM almost always provides a high detection rate and a low minimum false positives but with high computational demand". Their HMM-based technique gave lower detection rates, yet required lower number of training samples. By using $780,000$ system calls for training, the resulting detection rate was $97\%$.

A number of intrusion detection systems used sequences of system calls to train a Hidden Markov Model (HMM) classifier~\cite{forrest1999}\cite{wang2004}\cite{yeung2003}\cite{cho2003}\cite{hoang2003}. However, each system differs in the technique used for raising anomaly signal. Wang et al.~\cite{wang2004}, for example, raise anomaly signal when the probability of the whole sequence is below certain threshold. Warrender et al.~\cite{forrest1999}, on the other hand, declares a sequence as anomalous when the probability of one system call within a sequence is below the threshold. Cho and Park~\cite{cho2003} used HMM for modeling normal root privilege operations only. Hoang et al.~\cite{hoang2003} introduced a multi-layer detection technique that combines both outcomes from applying the Sliding Window approach and the HMM approach. 

Warrender et. al compared STIDE~\cite{forrest1996}, RIPPER~\cite{lee1998}, and HMM-based methods in~\cite{forrest1999}. They concluded that all methods performed adequately, while HMM gave the best accuracy on average. However, it required higher computational resources and storage space, since it makes multiple passes through the training data, and stores significant amount of intermediate data, which is computationally expensive, especially for large traces. 

The \textit{Kernel State Modeling} (KSM) technique represents traces of system calls as states of Kernel modules~\cite{murtaza2013}. The technique observes three critical states, namely Kernel (KL), Memory Management (MM), and File System (FS) states. The technique then detects anomaly by calculating the probability of occurrences of the three observed states in each trace of system calls, and comparing the calculated probabilities against the probabilities of normal traces. Applied to Linux-based programs of the UNM dataset, the KSM technique shows higher detection rates and lower false positive rates, compared to STIDE and HMM-based techniques.
\section{Conclusion and Future Work}
\label{sec_conclusion}
In this paper, we have shown the results of applying the BoSC technique for detecting anomaly in the behavior of Linux containers. We were able to show that malicious sequences created a strong anomaly signal, resulting in a $100\%$ detection rate, with low false positive rate of $0.58\%$. 

We used the BoSC technique to train the classifier, and test for anomalies in offline mode. Our next step is to modify the algorithm to be more suitable for deployment in a real-time intrusion detection system. In addition, some of the parameters currently used by the system are selected at random. Future work is to be directed to studying the effect of such parameters on the learning process and the detection speed and accuracy, in an attempt to formalize a way for optimizing their values for the target application. 
\ifCLASSOPTIONcompsoc
  \section*{Acknowledgments}
\else
  \section*{Acknowledgment}
\fi
This work was funded by Northrop Grumman Corporation via a partnership agreement through S2ERC; an NSF Industry/University Cooperative Research Center.  
We would like to express our appreciation to Donald Steiner and Joshua Shapiro for their support and collaboration efforts in this work.

\bibliography{references}

\begin{thebibliography}{10}
\providecommand{\url}[1]{#1}
\csname url@samestyle\endcsname
\providecommand{\newblock}{\relax}
\providecommand{\bibinfo}[2]{#2}
\providecommand{\BIBentrySTDinterwordspacing}{\spaceskip=0pt\relax}
\providecommand{\BIBentryALTinterwordstretchfactor}{4}
\providecommand{\BIBentryALTinterwordspacing}{\spaceskip=\fontdimen2\font plus
\BIBentryALTinterwordstretchfactor\fontdimen3\font minus
  \fontdimen4\font\relax}
\providecommand{\BIBforeignlanguage}[2]{{%
\expandafter\ifx\csname l@#1\endcsname\relax
\typeout{** WARNING: IEEEtran.bst: No hyphenation pattern has been}%
\typeout{** loaded for the language `#1'. Using the pattern for}%
\typeout{** the default language instead.}%
\else
\language=\csname l@#1\endcsname
\fi
#2}}
\providecommand{\BIBdecl}{\relax}
\BIBdecl

\bibitem{fuller2005}
D.~Fuller and V.~Honavar, ``{Learning classifiers for misuse and anomaly
  detection using a bag of system calls representation},'' in \emph{Proceedings
  of the Sixth Annual IEEE Systems, Man and Cybernetics (SMC) Information
  Assurance Workshop}.\hskip 1em plus 0.5em minus 0.4em\relax IEEE, 2005, pp.
  118--125.

\bibitem{mutz2006}
D.~Mutz, F.~Valeur, G.~Vigna, and C.~Kruegel, ``Anomalous system call
  detection,'' \emph{ACM Transactions on Information and System Security
  (TISSEC)}, vol.~9, no.~1, pp. 61--93, 2006.

\bibitem{alarifi2012}
S.~Alarifi and S.~Wolthusen, ``Detecting anomalies in {IaaS} environments
  through virtual machine host system call analysis,'' in \emph{International
  Conference for Internet Technology And Secured Transactions}.\hskip 1em plus
  0.5em minus 0.4em\relax IEEE, 2012, pp. 211--218.

\bibitem{alarifi2013}
------, ``Anomaly detection for ephemeral cloud {IaaS} virtual machines,'' in
  \emph{Network and system security}.\hskip 1em plus 0.5em minus 0.4em\relax
  Springer, 2013, pp. 321--335.

\bibitem{forrest1996}
S.~Forrest, S.~Hofmeyr, A.~Somayaji, and T.~Longstaff, ``A sense of self for
  unix processes,'' in \emph{Proceedings of the 1996 IEEE Symposium on Security
  and Privacy}, May 1996, pp. 120--128.

\bibitem{mysqlslap}
\BIBentryALTinterwordspacing
(2015) {mysqlslap - Load Emulation Client}. [Online]. Available:
  \url{http://dev.mysql.com/doc/refman/5.6/en/mysqlslap.html}
\BIBentrySTDinterwordspacing

\bibitem{sqlmap}
\BIBentryALTinterwordspacing
B.~Damele and M.~Stampar. (2015) {sqlmap: Automatic SQL injection and database
  takeover tool}. [Online]. Available: \url{http://sqlmap.org}
\BIBentrySTDinterwordspacing

\bibitem{forrest1999}
C.~Warrender, S.~Forrest, and B.~Pearlmutter, ``Detecting intrusions using
  system calls: alternative data models,'' in \emph{Security and Privacy, 1999.
  Proceedings of the 1999 IEEE Symposium on}, 1999, pp. 133--145.

\bibitem{hofmeyr1998}
S.~Hofmeyr, S.~Forrest, and A.~Somayaji, ``Intrusion detection using sequences
  of system calls,'' \emph{Journal of computer security}, vol.~6, no.~3, pp.
  151--180, 1998.

\bibitem{wang2004}
W.~Wang, X.-H. Guan, and X.-L. Zhang, ``Modeling program behaviors by hidden
  markov models for intrusion detection,'' in \emph{Machine Learning and
  Cybernetics, 2004. Proceedings of 2004 International Conference on},
  vol.~5.\hskip 1em plus 0.5em minus 0.4em\relax IEEE, 2004, pp. 2830--2835.

\bibitem{yeung2003}
D.-Y. Yeung and Y.~Ding, ``Host-based intrusion detection using dynamic and
  static behavioral models,'' \emph{Pattern recognition}, vol.~36, no.~1, pp.
  229--243, 2003.

\bibitem{cho2003}
S.-B. Cho and H.-J. Park, ``Efficient anomaly detection by modeling privilege
  flows using hidden markov model,'' \emph{Computers and Security}, vol.~22,
  no.~1, pp. 45 -- 55, 2003.

\bibitem{hoang2003}
X.~D. Hoang, J.~Hu, and P.~Bertok, ``A multi-layer model for anomaly intrusion
  detection using program sequences of system calls,'' in \emph{Proc. 11th IEEE
  Int?l Conf. Networks}, 2003, pp. 531--536.

\bibitem{lee1998}
W.~Lee and S.~J. Stolfo, ``Data mining approaches for intrusion detection,'' in
  \emph{Usenix Security}, 1998.

\bibitem{murtaza2013}
S.~S. Murtaza, W.~Khreich, A.~Hamou-Lhadj, and M.~Couture, ``A host-based
  anomaly detection approach by representing system calls as states of kernel
  modules,'' in \emph{Software Reliability Engineering (ISSRE), 2013 IEEE 24th
  International Symposium on}.\hskip 1em plus 0.5em minus 0.4em\relax IEEE,
  2013, pp. 431--440.

\end{thebibliography}
\end{document}